# Magnetic order and the electronic ground state in the pyrochlore iridate $Nd_2Ir_2O_7$

S. M. Disseler[1], Chetan Dhital[1], T. C. Hogan[1], A. Amato[2], S. R. Giblin[3], Clarina de la Cruz[4], A. Daoud-Aladine[3], Stephen D. Wilson[1], and M. J. Graf [1]*

[1] Department of Physics, Boston College, Chestnut Hill, MA 02467 USA
[2] Paul Scherrer Institute, CH 5232 Villigen PSI, Switzerland
[3] Rutherford Appleton Laboratory, Didcot, Oxfordshire OX11 0QX, UK
[4] Quantum Condensed Matter Division, Oak Ridge National Laboratory, Oak Ridge, TN 37831-6393, USA

Abstract

We report a combined muon spin relaxation/rotation, bulk magnetization, neutron scattering, and transport study of the electronic properties of the pyrochlore iridate $Nd_2Ir_2O_7$. We observe the onset of strongly hysteretic behavior in the temperature dependent magnetization below 120 K, and an abrupt increase in the temperature dependent resistivity below 8 K. Zero field muon spin relaxation measurements show that the hysteretic magnetization is driven by a transition to a magnetically disordered state, and that below 8 K a complex magnetically ordered ground state sets in, as evidenced by the onset of heavily damped spontaneous muon precession. Our measurements point toward the absence of a true metal-to-insulator phase transition in this material and suggest that $Nd_2Ir_2O_7$ lies either within or on the metallic side of the boundary of the Dirac semimetal regime within its topological phase diagram.

Recently, correlated iridium oxide compounds have attracted a great deal of interest due to the delicate interplay between electron-electron correlation effects and spin-orbit induced band renormalization manifest in these materials [1-3]. As a result, a number of novel topological phase transitions have been proposed as the correlations in these systems are tuned from strongly correlated Mott insulators to weakly correlated topological band insulators—the most striking of which is the Dirac semimetal phase. This topological phase is predicted to possess topologically protected electronic states along only particular momentum points and crystallographic surfaces in a three-dimensional solid with magnetic correlations [2]. The most promising iridates for exploring this topological phase diagram are currently the $A_2Ir_2O_7$ (A-227) pyrochlore compounds where, through tuning the A-site, the system transitions from a magnetically ordered insulator into a spin disordered, unconventional metal [4-9].

Specifically, near the transition from the Mott insulating phase to the metallic ground state within the topological phase diagram, the topologically nontrivial Dirac semimetallic phase is predicted to emerge [2]. In tuning the A-site ionic radius and the corresponding bandwidth of the A-227 iridates, such a transition likely occurs between the known correlated insulating phase of $Eu_2Ir_2O_7$ [6,7] and the unconventional metallic phase of $Pr_2Ir_2O_7$ [9]. The insulating variants of the A-227 series all display metal-to-insulator (MI) phase transitions at their respective $Ir^{4+}$ ordering temperatures; however a number of recent studies have suggested that $Nd_2Ir_2O_7$ (Nd-227) shows a dramatically reduced MI transition onset temperature [10,11]. This supports a developing picture of Nd-227 residing just inside the correlated, insulating boundary of the Dirac semimetal phase transition line. Studies of the intrinsic behavior of the Nd-227 ground state are thus an important metric for exploring the viability of the proposed topological phase diagram in the pyrochlore iridates and for exploring Nd-227's capacity for being tuned into the Dirac semimetal phase.

A number of conflicting experimental reports have revealed a rather complex picture of the intrinsic Nd-227 electronic ground state. These studies have suggested that the reported insulating phase in this material is a delicate balance between sample stoichiometry and other extrinsic factors, and the precarious nature of the insulating state is evidenced by the variability in previously observed properties: initial reports by Yanagishima et al. showed Nd-227 was clearly metallic at all temperatures [6], whereas in later studies by Matsuhira et al. a MI transition was reported at 37 K [10]. Subsequent studies of the resistivity under pressure [11] revealed a small upturn in the resistivity below 20 K - interpreted as the onset of the MI transition - which was suppressed under an increased pressure of 10 GPa. At these higher pressures, a small resistive anomaly was observed near $T = 3.3$ K that was associated with the onset of magnetic order via an RKKY interaction facilitated by the suppression of the MI transition.

The magnetic order associated with this upturn in resistivity was initially presumed to be ferromagnetic based on the detailed magnetic response of the resistivity, with a 2-in/2-out structure of the $Nd^{3+}$ ($J = 9/2$) moments. Recent neutron scattering measurements, however, suggest that magnetic order occurs at ambient pressure with long-range magnetic order (LRMO) of the $Ir^{4+}$ moments setting in below 15 K [12] in a structure consistent with an all-in/all-out arrangement. The large decoupling between the reported MI transition temperature (T ~ 37K) and onset of correlated magnetism is anomalous and strongly suggestive of additional interactions relevant in the ground state of this material.

In order to explore the interactions between the correlated spin order and the charge carriers in this material, in this letter we present a combined magnetization, transport, neutron diffraction, and muon spin relaxation/rotation (μSR) study on polycrystalline samples of Nd-227. By leveraging both bulk and local probes on the same sample, we find the onset of a disordered magnetic state at 120 K associated with the $Ir^{4+}$ sublattice, similar to that observed for other members of the A-227. Upon continued cooling below 8 K, we observe the onset of long-range magnetic order likely associated with the $Ir^{4+}$ sublattice. Surprisingly however, our combined results find no evidence for a metal-to-insulator transition in this system at any temperature, suggesting that Nd-227 system may already lie within the Dirac semimetal regime of the topological phase diagram.

Polycrystalline samples of $Nd_2Ir_2O_7$ were synthesized by reacting stoichiometric amounts of $Nd_2O_3$ (99.99%) and $IrO_2$ (99.9%). Powders were pelletized using an isostatic cold press and reacted at temperatures between 900 C to 1125 C over a period of six days with several intermediate grindings. The samples were determined to be phase-pure with the exception of two minor impurity phases of $IrO_2$ and $Nd_2O_3$ each comprising less than 1% of the total volume fraction. Neutron experiments were performed on the HB-2A powder diffractometer at the High Flux Isotope Reactor at Oak Ridge National Lab and on the HRPD instrument at the ISIS spallation neutron facility. Magnetization measurements were performed in an Oxford MagLab dc-extraction magnetometer. The electrical resistivity was measured using standard AC four-probe techniques in gas flow and $^3$He cryostats with a 9T magnet. Magnetic fields were applied parallel to the current, and the excitation level was varied at several temperatures to ensure no self-heating occurred. Muon spin relaxation (µSR) measurements were carried out over the temperature range 1.6 K < $T$ < 150 K on the EMU spectrometer at the ISIS pulsed beam facility at the Rutherford Appleton Laboratories, and the GPS spectrometer on the πM3 continuous beamline at Paul Scherrer Institute (PSI). Powder samples were studied at ISIS in a silver holder with a Mylar window, and the small contribution (10%) of muons stopping in the silver was independently measured and subtracted from the data. Powder samples at PSI were sealed in metalized Mylar packets, with no background contribution.

The temperature dependent static susceptibility χ($T$)=$M/H$ of polycrystalline Nd-227 was measured from 250 K to 1.8 K in samples under a field of 1000 Oe under both zero field cooled (ZFC) and field cooled (FC) conditions, and the results are shown in Fig. 1a. Under ZFC conditions, we observe a small peak in of Nd-227 at 105 K, while under FC conditions we see a sharp increase at $T$ = 120 K. This behavior is similar to that observed for other members of the A-227 family and occurs at a comparable temperature [6,7]. It is associated with the onset of magnetism in the $Ir^{4+}$ sublattice, as it occurs when the A-site species is either magnetic or non-magnetic. The anomaly is typically linked to a MI transition, which occurs at comparable temperatures. We note however that the onset temperature of the magnetic anomaly from these magnetization measurements is considerably higher than the value of $T_{MI}$ = 37 K extracted from earlier resistivity measurements [10]. Upon continued cooling below 100 K there is a significant difference between the FC and ZFC behavior until 8 K, below which the ZFC and FC curves converge and continue to increase weakly with a further decrease in temperature.

Turning to resistivity measurements on the same sample, the temperature dependent resistivity of Nd-227 is shown in Fig, 1b. A decrease in the resistivity is observed below room temperature and reaches a broad minimum at T ~ 65 K. Between 65 K and 8 K there is a very weak increase in the resistivity; however below 8 K there is a second slope-change followed by a slow, nearly logarithmic increase in resistance with decreasing temperature (inset of Fig. 1b) below 1 K. This logarithmic temperature dependence differs greatly from the exponential behavior expected for an insulating state, but is characteristic of correlations in a weakly metallic system (e.g., Kondo effect); this strongly suggests the absence of a metal-insulator transition in this system. An applied magnetic field of 9 T in the region above 8 K has very little effect on the resistance, in contrast to the large negative magnetoresistance that is observed at lower temperatures. To verify this behavior, samples from two different batches were measured, including one sample before and after a 24 hour anneal in an oxygen environment, and in all cases the data were the same, demonstrating that the observed behavior is insensitive to small changes in chemical composition.

In Fig. 2, we show the magnetic field dependence at $T = 1.8$ K of both the magnetization and resistivity. The $J = 9/2$ state is known to split into 5 Kramers doublets in the crystal field of the magnetic ion, assuming a trigonal symmetry [13]. Therefore, a nonlinear least-squares fit of a Brillouin function with $J_{eff} = ½$ was used to describe the data. The fit is shown as the dashed curve in Fig. 2, corresponding to paramagnetic moments with $J_{eff} = ½$ and an effective moment $\mu_{eff} = 1.3 \pm 0.1$ $\mu_B$. This $\mu_{eff}$ is larger than the maximum estimated size of the $Ir^{4+}$ moment but smaller than the free $Nd^{3+}$ moment. From this, we conclude that the Nd moments remain paramagnetic down to at least 1.8 K whereas the $Ir^{4+}$ moments undergo a magnetic transition at 120 K.

As an initial probe of the correlated $Ir^{4+}$ spin order, elastic neutron scattering measurements were performed. The resulting neutron diffraction patterns were refined using the Fd-3m cubic space group, and the lattice parameters were determined to be $a =$ 10.3588(4) Å at 4 K and $a =$ 10.3647(9) Å at 200 K. However, our measurements did not show any evidence of long-range magnetic order down to 4 K with the upper bound of an ordered moment conservatively estimated to be 0.5 $\mu_B$. We also performed high-resolution time-of-flight diffraction measurements on HRPD, at ISIS on this same sample in order to improve the statistical significance of this result. From these combined measurements, we can conclude that within comparable statistical confidence of Ref. 12 that the long-range magnetic signal reported below 15 K in earlier reports is not present in our sample [14].

In order to explore the magnetic order further, µSR measurements were performed on two spectrometers with complementary resolutions/sensitivities to both fast relaxation processes (GPS at PSI) and slow relaxation processes (EMU at ISIS). By combining these measurements, we obtain a picture of both static and fluctuating magnetic fields throughout the bulk of the sample over a broad frequency domain. In modeling the data at higher temperatures ($T \geq 8$ K), the time-dependent depolarization curves can be fit by a stretched exponential function

$$P(t) = A_S \exp\left[-(\lambda_S t)^\beta\right], \qquad (1)$$

where $A_S$ is the asymmetry of the depolarization, normalized to the independently measured full asymmetry, $\lambda_S$ is the slow depolarization rate and β is the stretched exponent. For the ISIS data, $A_S$ and $\lambda_S$ are both left as fit parameters, since the onset of fast depolarization will be manifest in a loss of apparent asymmetry. For the GPS data, the asymmetry was independently measured and $A_S$ was fixed at this value. The results are shown in Fig. 3 where above 10 K we see excellent agreement between the results taken at the two different facilities. The slow depolarization rate $\lambda_S$ shows a clear change at 120 K, close to the temperature at which the magnetization exhibits the onset of hysteretic behavior (see inset of Fig. 3). Below 10 K, a dramatic increase in the relaxation rate is observed. For the EMU data, $A_S$ is nearly constant near its full value at higher temperatures indicating that fast relaxation processes are negligible. However, below about 10 K $A_S$ drops rapidly, reaching a value of approximately 0.39(1) of the full asymmetry at 1.6 K. This behavior clearly demonstrates the onset of fast depolarization processes, and accounts for the difference in $\lambda_S$ for the two data sets below 10 K. The exponent β (results not shown) for both the GPS and EMU data has a value $1.05 \pm 0.02$ at high temperatures, drops to $0.80 \pm 0.02$ below 120 K, then drops again below 20 K to a value of $0.50 \pm 0.05$ at 8 K.

In Fig. 4, we show a sequence of low-temperature depolarization curves taken on GPS, where the fast relaxation processes can be observed. Upon cooling below $T = 8$ K, we see the onset of spontaneous oscillations, unambiguously demonstrating the existence of magnetic ordering. The oscillations are heavily damped, and attempts to fit the data to the expected two-component depolarization function for polycrystalline samples with magnetic order as utilized in Ref. 4 and 5 were unsuccessful. A three-component depolarization function of the form

$$P(t) = A_1 \exp(-\lambda_1 t)\cos(\omega_\mu t + \phi) + A_2 \exp(-\lambda_2 t) + A_S \left[-(\lambda_S t)^\beta\right], \quad (2)$$

yields an adequate fit to the data, as shown in Fig. 4. We find a muon precession frequency $\omega_\mu/2\pi$ = 8.8(2) MHz, corresponding to an average local field for the precessing muons of $\langle B_{loc} \rangle = \omega_\mu/\gamma_\mu$ = 665 G, where $\gamma_\mu$ is the muon gyromagnetic ratio and $\gamma_\mu/2\pi$ = 0.01355 MHz/G. This value is consistent with Fourier transforms of the data, although the Fourier peak is quite broad, with a half-width of approximately 5 MHz. The extracted frequency is significantly smaller than the 13.3 MHz value observed for $Eu_2Ir_2O_7$ ($\langle B_{loc} \rangle$ = 987 G) [4], and the 14.8 MHz ($\langle B_{loc} \rangle$ = 1100 G) found for $Y_2Ir_2O_7$, and $Yb_2Ir_2O_7$ [5]. The difference in frequency is apparent when comparing the $T$ = 1.8 K depolarization curves for Nd-227 and Yb-227 [5] plotted in Fig. 4.

The depolarization rates are $\lambda_1$ = 13(1) $\mu s^{-1}$, $\lambda_2$ = 15(2) $\mu s^{-1}$, $\lambda_s$ = 0.18(1) $\mu s^{-1}$, and $\beta$ = 1.07(6); the relative amplitudes are $A_1$ = 9.3, $A_2$ = 6.7(6), and $A_s$ = 7.1(4); the total amplitude $A_1 + A_2 + A_s$ was constrained to be the measured full asymmetry, with the ratio of slowly decaying asymmetry to the total asymmetry, $\eta \equiv A_s/A_t$, found to be approximately 0.30(2). The resultant phase angle is extremely large, $\phi$ = -63° and the fit deviates from the data at very short times.

We also achieved a reasonable alternative fit of the data at 1.6 K using a two-component Bessel function

$$P(t) = A_1 \exp(-\lambda_1 t) J_0(\omega_\mu t) + A_S \left[-(\lambda_S t)^\beta\right], (3)$$

where $J_0$ is the spherical Bessel function of the first kind, as commonly used in systems exhibiting spin density wave ordering [15]. For Eq. 3, we extracted the parameters $\omega_\mu/2\pi$ = 8.26 MHz, $\lambda_1$ = 9.8 $\mu s^{-1}$, $\lambda_1$ = 0.23 $\mu s^{-1}$ $\beta$=0.45, and $\eta$ = 0.42. Although this fit captured the very short and long times accurately, it failed to accurately describe the entire time interval below 1 $\mu$s. Nonetheless, both fits allow us to reasonably estimate the local field, and extract the rate of the slow depolarization term below 8 K (also shown in Fig. 3 for Eq. 2). Here the relative contribution of $A_S$ of the total asymmetry and $\lambda_s$ are found to be roughly independent of temperature below 5 K. Longitudinal field studies (see Supplemental Information [14]) show that the slow relaxing component is unaffected by fields up to 300 G. Combined with the temperature independence of the depolarization rate below 5 K (Fig. 3), this suggests that quantum rather than thermal fluctuations cause the slow depolarization.

We now turn to the nature of the low-temperature ordered magnetic state. Our magnetization results strongly suggest that the ordering is on the $Ir^{4+}$ sublattice, while the $Nd^{3+}$ moments remain in a paramagnetic state. This ordered state is unusual, as evidenced by the highly damped oscillations and lack of detectable neutron scattering peaks associated with this order. Assuming the muon stopping site(s) are the same for all the A-227 materials, the low muon precession frequency observed for Nd-227 indicates that the magnetic structure is different than that for the Eu, Y, and Yb based materials. We infer from these results two different scenarios: the underlying correlations may produce a 'small-moment' system, as observed in $UPt_3$ or $URu_2Si_2$ [16], or alternatively, the magnetic order could have a small correlation length. The muon depolarization following Eq. 3, based on a Bessel function, favors the former scenario, while the three-component depolarization described by Eq. 2 combined with the potential presence of strong frustration in the system would seem to favor the latter scenario.

The onset of magnetism at 120 K, as indicated by both the magnetization and μSR measurements, combined with the lack of spontaneous muon precession at this temperature indicates there is no long-range ordering at this transition. In this temperature region, the system is clearly metallic, demonstrating that geometric frustration, if present, is not relieved by an MI transition, and these results are in contrast to the ordering observed around this same temperature in the insulating Eu, Y, and Yb based materials. This ordered phase may be of the spin-glass variety, and our data are also consistent with the prediction of a highly degenerate 2in/2out structure [17]. The resulting behavior is intermediate between the results reported for $Pr_2Ir_2O_7$ (disordered metallic ground state [8,9]) and the three insulating systems mentioned above, confirming the variation of correlations and change in the band structure with ionic radius.

Our data fail to show that a MI transition occurs in this system, as evidenced by the logarithmic temperature dependence of the resistivity at low temperatures. Also, the variation of the magnetoresistance $\Delta\rho=\rho(H)/\rho(0)-1$ with magnetization is *negative* and varies linearly with $M^2$ (inset, Fig 2), analogous to the Kondo screening observed in $Pr_2Ir_2O_7$, where both the RKKY interaction strength and Kondo temperature are approximately 20 K [9]. At ambient pressure, no magnetoresistance has been reported in A-227 with a nonmagnetic A-site; however removing the MI transition of $Eu_2Ir_2O_7$ through the application of high pressure [18] showed that this material exhibits positive magnetoresistance proportional to $H^2$, similar to normal metals. We assert that an additional magnetic scattering channel must be present in Nd-227, and it is this process, rather than a MI transition that causes the upturn in the resistivity below 8K. Hence, we propose that Nd-227 exhibits a magnetically ordered, metallic ground state: it is weakly metallic, with magnetic order facilitated via RKKY interactions.

In summary, we have studied high quality $Nd_2Ir_2O_7$ samples via magnetization, resistivity, neutron diffraction, and μSR. We find the onset of disordered magnetism at $T$ = 120 K, likely arising from the geometric frustration of the $Ir^{4+}$ sublattice. Below $T$ = 8 K, magnetic order appears albeit heavily damped; however we find no evidence for a metal-insulator transition, suggesting that the low temperature magnetic ordering is facilitated via the RKKY interaction. While more detailed studies via both μSR and neutron scattering are required to understand the structure of the magnetically ordered state, our current data demonstrate that the $Nd_2Ir_2O_7$ compound lies just within the metallic phase boundary, potentially within the Dirac semi-metal regime, of the topological phase diagram of the pyrochlore iridates [2].

M.J.G. and S.D.W. would like to acknowledge very helpful discussions with Ying Ran. This work was supported in part by National Science Foundation Materials World Network grant DMR-0710525 (M.J.G.) and by NSF CAREER award DMR-1056625 (S.D.W.). Muon experiments were performed at the ISIS Muon Facility at the Rutherford Appleton Laboratories (UK) and the Swiss Muon Source at the Paul Scherrer Institute (Switzerland). Part of this work was performed at Oak Ridge National Laboratory High Flux Isotope Reactor, sponsored by the Scientific User Facilities Division, Office of Basic Energy Sciences, U.S. Department of Energy.

**Figure Captions**

Figure 1. (a) Magnetization versus temperature in an applied field of 1000 G for field cooled (solid symbols) and zero field cooled (open symbols) sample of $Nd_2Ir_2O_7$. Inset: Expanded view of the susceptibility in the vicinity of $T = 120$ K. (b) Resistivity versus temperature for $Nd_2Ir_2O_7$ in applied fields of 0 T, 4 T, and 9 T. Inset: Temperature dependent resistivity at low temperatures in zero applied field.

Figure 2. Magnetic field dependence of the resistivity and magnetization at $T = 1.8$ K. The dashed line shows the fit to a Brillouin function as described in the text. Inset: variation of the fractional change in resistivity with the square of the magnetization, with magnetic field as an implicit parameter; the straight line is a guide to the eye.

Figure 3. Temperature dependent muon depolarization rate for extracted from data taken at ISIS and PSI, as described in the text. Inset: Expanded view of the depolarization rate in the vicinity of $T = 120$ K. Solid lines are guides to the eye.

Figure 4. Evolution of the short-time muon depolarization curves (PSI) with decreasing temperature. The upper four curves are for $Nd_2Ir_2O_7$, at temperatures of 8 K, 5 K, 3.5 K, and 1.6 K from top to bottom. The bottom curve is for $Yb_2Ir_2O_7$ at 1.6 K (from Ref. 10). Curves are offset for clarity. The solid lines are fits to Eq. 1 for the $T = 8$ K data and to Eq. 2 for the $T = 1.6$ K data for the $Nd_2Ir_2O_7$ curves; the solid line for the Yb curve is a fit to the depolarization function as described in Ref. 9.

Figure 1

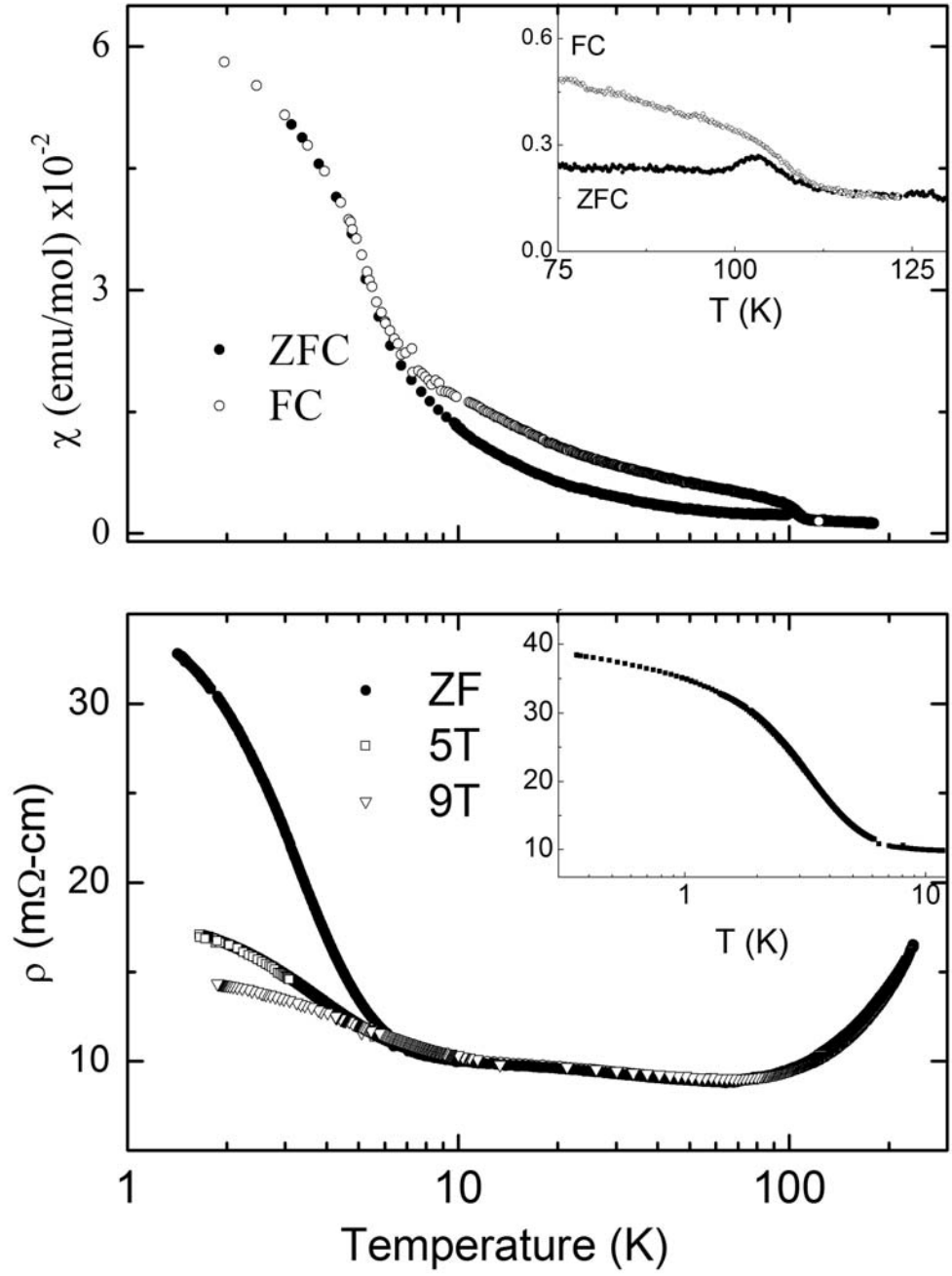

Figure 2

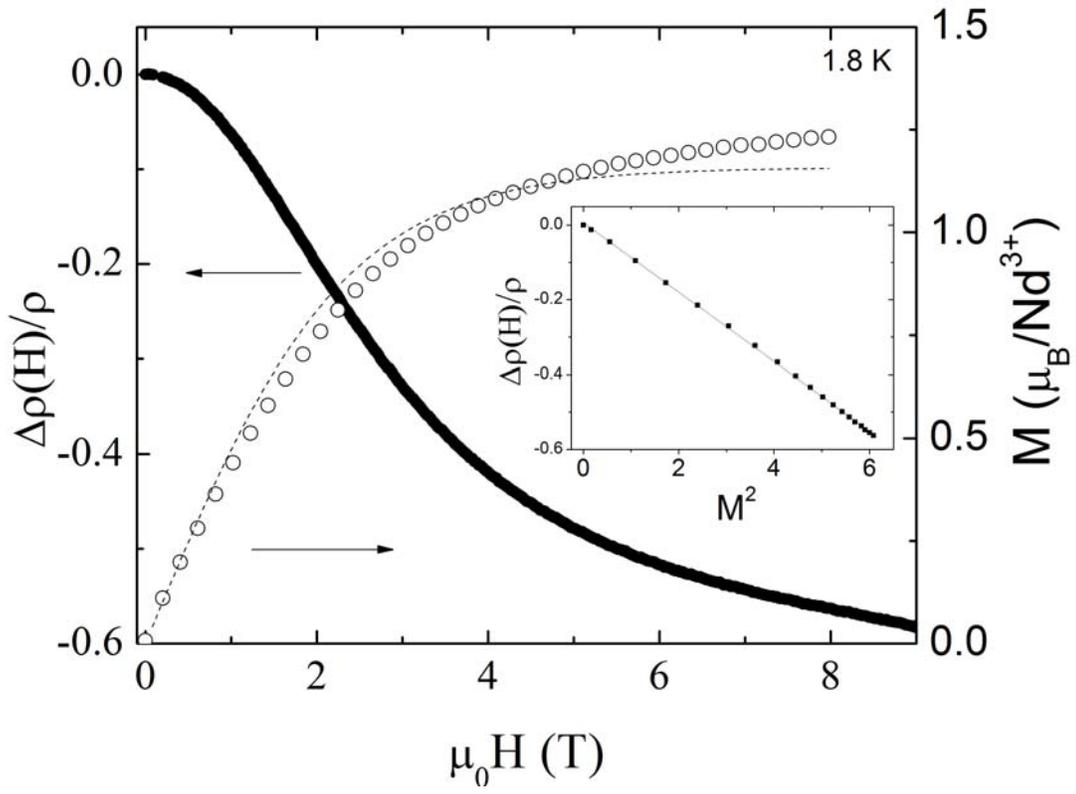

Figure 3

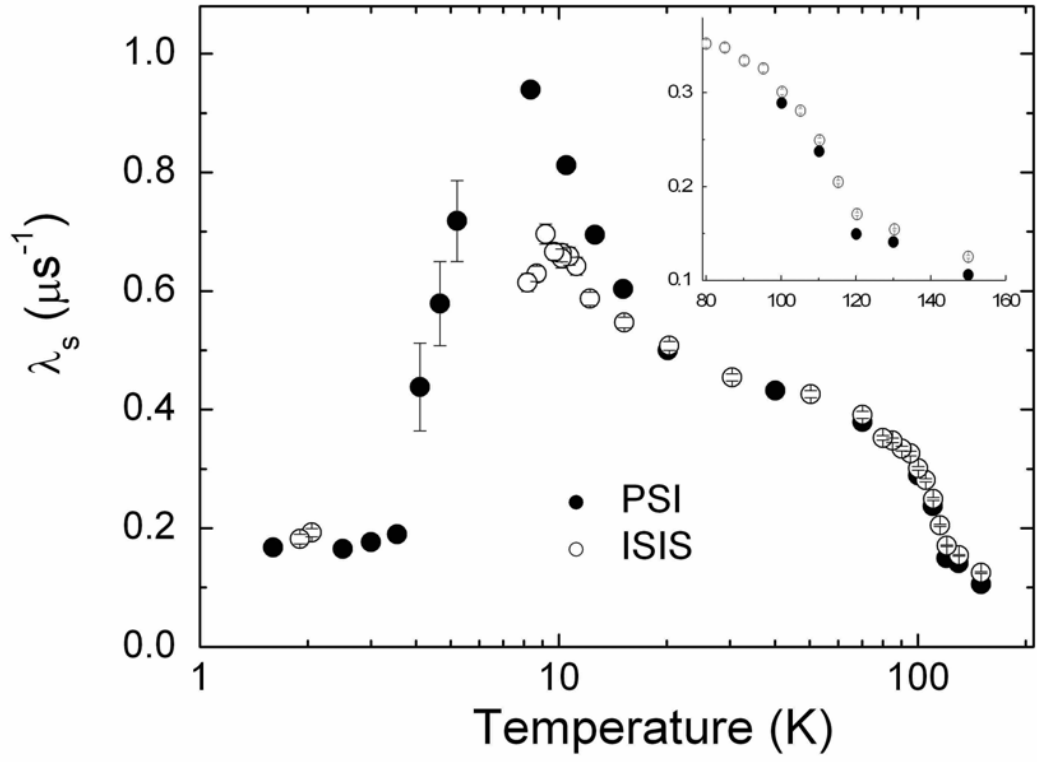

Figure 4

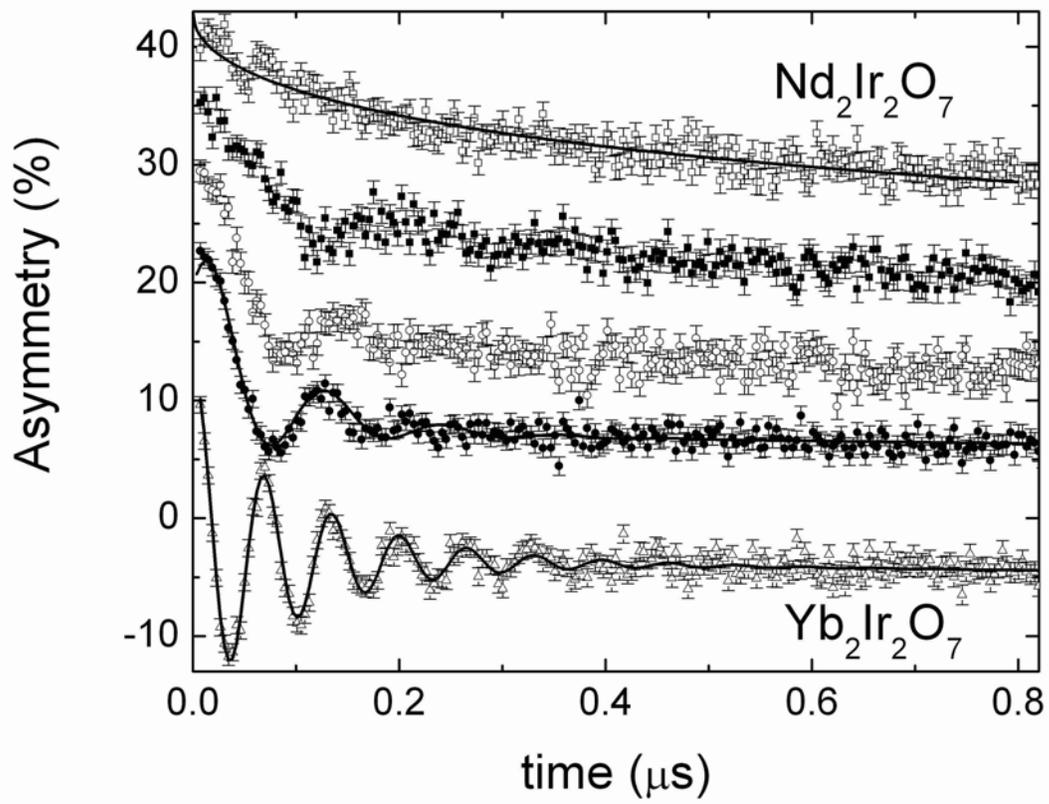

Magnetic order and the electronic ground state in the pyrochlore iridate

$Nd_2Ir_2O_7$

S. M. Disseler *et al.*

**Neutron diffraction studies:** Two separate batches of polycrystalline $Nd_2Ir_2O_7$ powder were studied on the HB-2A and HRPD diffractometers at the ORNL High Flux Isotope Reactor and the ISIS spallation neutron sources respectively. Data from HB-2A were collected with $\lambda_i$=1.5385 Å with a Ge(115) monochromator and 12'-31'-6' collimation, and data were refined using the FullProf software package. The results of the refinement at T=3K are plotted below in Fig. S1. Neither new antiferromagnetic Bragg peaks nor any ferromagnetic enhancement of nuclear peaks was observed upon cooling below the magnetic transitions at both 120K and 8K. This sets a conservative estimate of the maximum ordered moment of ~ 0.5 $\mu_B$. A highly disordered magnetic state with a small correlation length may also preclude detection of correlated magnetism in this measurement.

In order to increase the statistics of this initial measurement, we performed further diffraction measurements at the HRPD diffractometer at the ISIS spallation source. These additional measurements also showed no sign of magnetic order below the transitions at 120 K and 8 K within the resolution of the spectrometer. In order to directly compare with the order parameter reported by Tomiyasu *et al.* [12], we plot the integrated intensities from the Q = (2, 2, 2) and Q = (1, 1, 3) nuclear reflections at $T$ = 4 K, 75 K, 180 K, and 300 K (Fig. S2). The window of integration is plotted in Fig. S2 (a), and upon cooling from 300 K, we observe a continuous decrease in this integrated signal with no observable increase at 4K due to $Nd^{3+}$-ordering (Fig. S2 (b)). This integrated time-of-flight data approaches the statistical accuracy of the earlier report by Tomiyasu *et al.* and allows us to conclude that the ordered moment reported in $Nd_2Ir_2O_7$ is either substantially diminished or not present in our samples. This discrepancy may be due to an extreme sensitivity to the sample stoichiometry or oxygen content; however it is surprising given the higher energy scale of the $Ir^{4+}$ ordering anomaly observed at $T_{Ir}$ =

120K in our samples (by both μSR, magnetization, and resistivity) relative to the earlier reported $T_{Ir}$ = 36 K [7].

**Longitudinal field μSR studies:**

For the low temperature longitudinal field measurements conducted on EMU, the fast decaying component is not detected by the spectrometer and so at low temperatures we are only measuring the slowly relaxing component (equivalent to the third component in Eq. 2). Our results (see Fig. S3) at $T$ = 1.6 K for the fractional asymmetry $\eta_S = A_S/A_{total}$ and the depolarization rate $\lambda_S$ (via Eq. 1) have zero-field limits of $\eta_S$ = 0.36(1) and $\lambda_S$ = 0.17(2) μs$^{-1}$, in good agreement with the GPS results. A field of about 350 G is required to restore half of the missing asymmetry by decoupling the muon from the local magnetic field, roughly a factor two lower than the characteristic precession field of 665 G. We also find that the depolarization rate $\lambda_S$ is independent of applied field up to 300 G, at which point the muon is decoupling from the local field. Since $\lambda_S / \gamma_\mu \approx 2$ G, the lack of magnetic field dependence in $\lambda_S$ demonstrates that the long-time depolarization is dynamical in origin. The temperature independence of the depolarization rate below 5 K (Fig. 3 in the text) suggests that quantum rather than thermal fluctuations are important in this regime.

Supplemental Figure S1

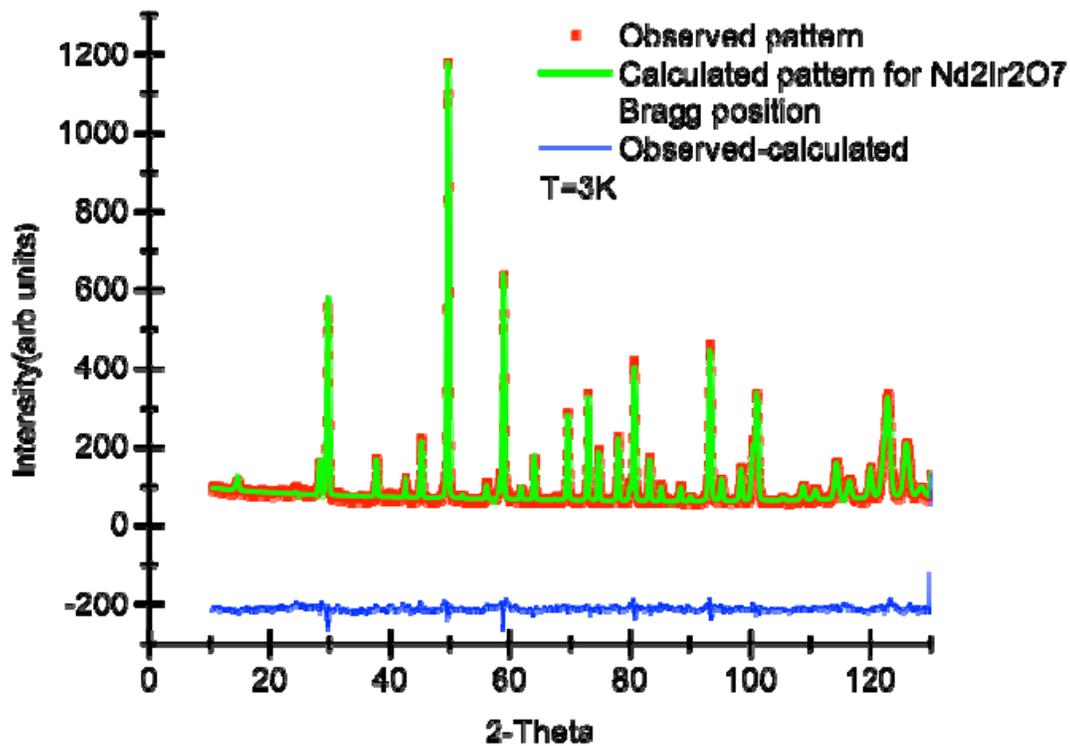

Supplemental Figure S1. Neutron powder diffraction data collected on HB-2A. The measured data (red circles) are plotted with the calculated pattern (green line) described in the text. Vertical tick marks denote the calculated Bragg positions and the solid blue line denotes the difference between the calculated and observed patterns.

Supplement Figure S2

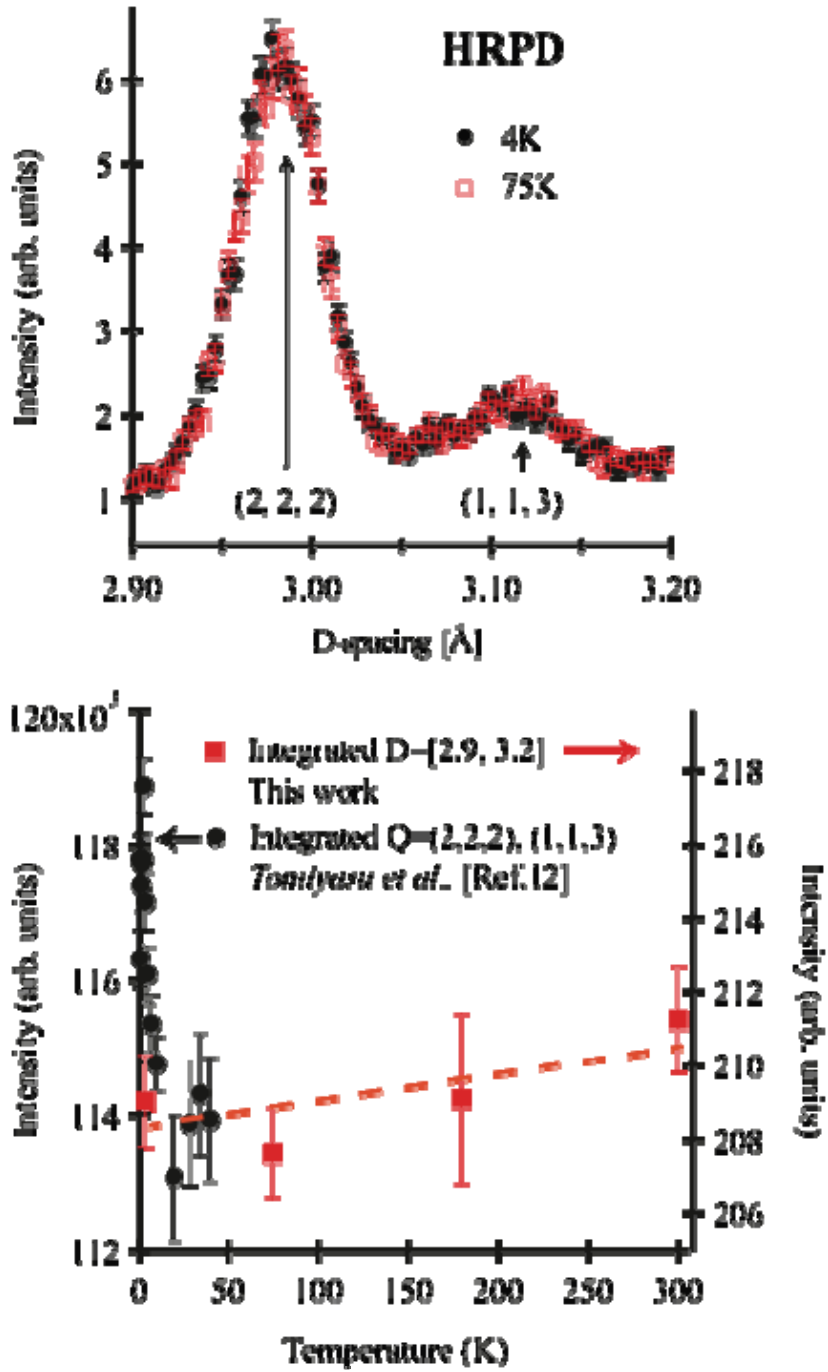

Supplemental Figure S2. Top panel shows scattering intensities measured at 4K (black circles) and 75K (open squares) in the low angle bank of HRPD. Bottom panel shows the integrated intensities measured between D-spacings of 2.9 and 3.2 Å as a function of temperature (red squares). Black circles are data taken from Ref. 12 for relative, statistical comparison.

Supplemental Figure S3

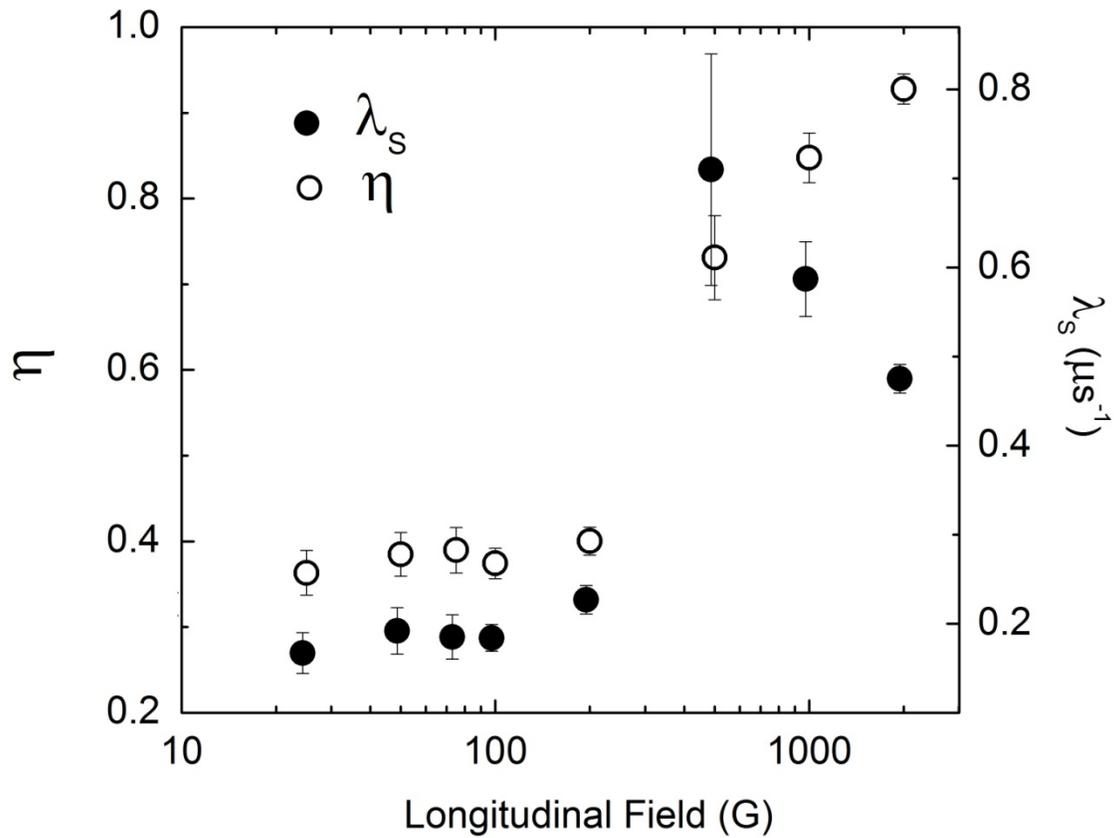

Supplemental Figure S3. Longitudinal magnetic field dependence of the slow depolarization rate $\lambda_S$ and its fractional asymmetry component $\eta$ at $T = 1.6$ K extracted from data taken at ISIS.